\documentclass[proceedings, preprint]{rmaa}

\usepackage{paralist}

\SetYear{2017}
\SetConfTitle{4th Robotic Telescope Workshop}
\SetVolume{ 5000000 }
\SetFirstPage{ 1 }
\ReceivedDate{ February 5th, 2018 }
\AcceptedDate{ as soon as possible }

\title{From a computer controlled telescope to a robotic observatory: the history of the VIRT} 

\author{
  B. Gendre,\altaffilmark{1,2} 
N.~B. Orange,\altaffilmark{2,3} 
D.~C. Morris,\altaffilmark{1,2} 
T. Giblin,\altaffilmark{4} 
J. Neff,\altaffilmark{5} 
A. Klotz,\altaffilmark{6} 
and P. Thierry\altaffilmark{7}}

\altaffiltext{1}{University of the Virgin Islands,
  College of Science and Math, 2 John Brewer's Bay, St Thomas, 00802, VI, USA
	(bruce.gendre@gmail.com).}

\altaffiltext{2}{Etelman Observatory, St Thomas, 00803, VI, USA.}

\altaffiltext{3}{OrangeWave Innovative Science, LLC, Moncks Corner, SC 29461, USA.}
\altaffiltext{4}{Department of Physics, United States Air Force Academy, Colorado Springs, CO 80840.}
\altaffiltext{5}{National Science Foundation, USA.}
\altaffiltext{6}{Universit\'e de Toulouse; UPS-OMP; IRAP; Toulouse, France ; CNRS; IRAP; 14, Avenue Edouard Belin, F-31400 Toulouse, France.}
\altaffiltext{7}{Auragne Observatory, France.}

\shortauthor{Gendre et al.}
\shorttitle{History of the VIRT}

\listofauthors{B. Gendre, N.~B. Orange, et al.}
\indexauthor{Gendre, B.}
\indexauthor{Orange, N.~B.}
\indexauthor{Morris, D.~C.}
\indexauthor{Giblin, T.}
\indexauthor{Neff, J.}
\indexauthor{Klotz, A.}
\indexauthor{Thierry, P.}

\abstract{The Virgin Island Robotic Telescope is located at the Etelman Observatory, St Thomas, since 2002. We will present its evolution since that date with the changes we have performed in order to modify an automated instrument, needing human supervision, to a fully robotic observatory. The system is based on ROS (Robotic Observatory Software) developed for TAROT and now installed on various observatories across the world (Calern, La Silla, Zadko, Les Markes, Etelman Observatory).}

\resumen{El Telescopio Rob\'otico Virgin Island est\'a ubicado en el Obervatorio
Etelman, St Thomas, desde 2002. Presentamos su evoluci\'on desde esa fecha con los cambios que hemos ido realizando para modificar un instrumento automatizado que requiere supervisi\'on humana a un obesrvatorio completamente robotizado. El sistema se basa en ROS (Observatorio de Software Rob\'otico) desarrollado para TAROT y ahora instalado en varios observatorios de todo el
mundo (i.e. observatorios de Calerno, La Silla, Zadko, Les Markes y de Etelman). }

\addkeyword{instrumentation: miscellaneous}
\addkeyword{techniques: miscellaneous}
\addkeyword{telescopes}

\begin{document}
% Typeset article header
\maketitle

\section{Introduction}
\label{intro}

The Etelman Observatory \citep{nef04} is an university owned facility located at the top of Crown Mountain, St Thomas, in the US Virgin Islands (as part of the Leeward islands). The building, a former house, was donated by the Etelman family in the 1970's to the (then) College of the Virgin Islands (now the University of the Virgin Islands, or UVI). H. Etelman was an amateur astronomer, and built a small dome on the roof of his house equipped with a 0.4m telescope on an independent pier. The donation stipulated that the house remained an active observatory to perform University research. This condition was accepted and for decades the observatory was used by the students of the College and the University.

The observatory was badly damaged by hurricane Marilyn in 1995, but was offering very interesting potential for a modern, automated, observatory. Following this idea, D. Drost, a former professor from UVI, proposed to refurbish the observatory with a new telescope and instrumentation. This project has been funded by various grants, and is ongoing since 1999. In this paper, we present the final modifications made to the observatory to allow for a complete robotization of the site. A companion paper (Morris et al., these proceedings) will expose the first results obtained to date.

We describe the instrument and the Etelman Observatory in Section \ref{tel}, set the scientific objectives in Section \ref{objectif}, before describing the software and hardware modifications needed to reach these objectives in Sections \ref{soft} and \ref{hard} respectively. We shortly tease the first results in Section \ref{result}, before concluding.

\section{The site and the instrument}
\label{tel}

\begin{figure*}[!t]\centering
  \vspace{0pt}
  \includegraphics[width=1.9\columnwidth]{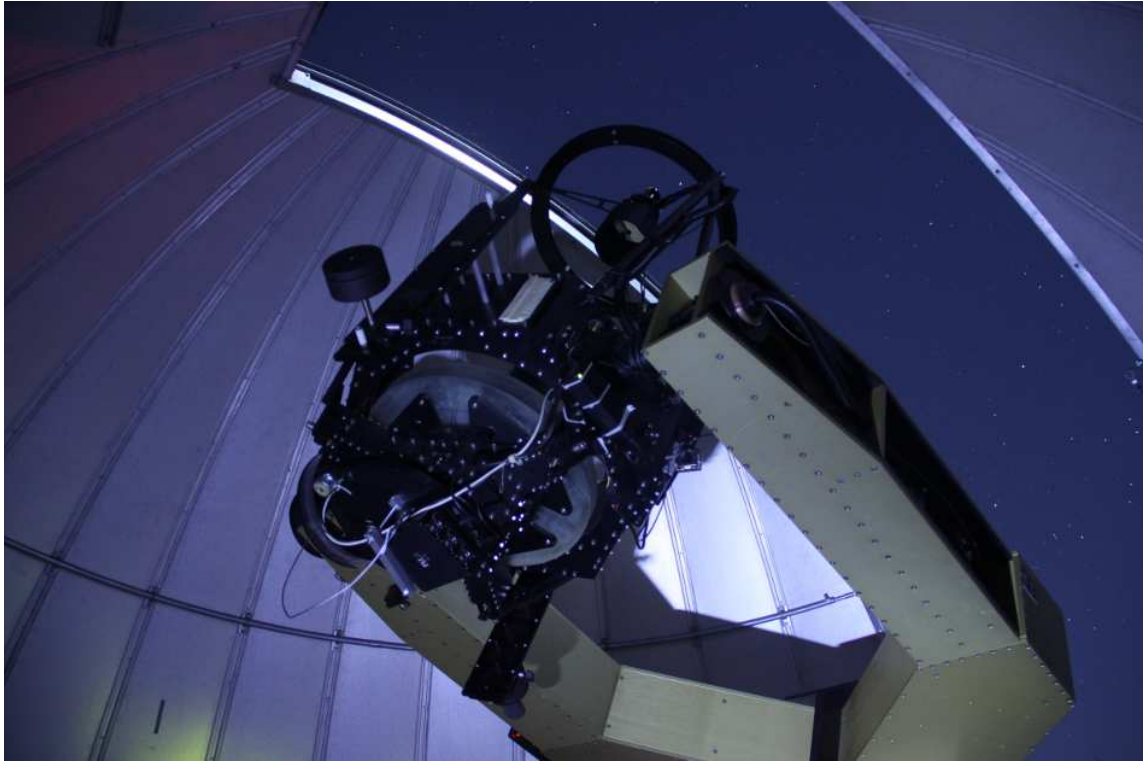}
  \caption{The Virgin Island Robotic Telescope.}
  \label{fig1}
\end{figure*}

The island of St. Thomas is located close to the equator, offering a view of the sky between $-70^o < \delta < +90^o$, and at about $65^o$ West in longitude, placing it as the most eastern facility in the United State. The observatory site is situated near the crest of Crown Mountain, offering a completely free view all around the telescope with most parts of the horizon constituted by the ocean. All of this allows for an efficient bridge between European and American northern observatories for a complete coverage of any phenomenon. The facility offers an existing infrastructure (power, water, roads, guest housing) allowing for operations during the whole year, albeit the hurricane season (June-October) may drive a few weeks off in case of storms. The sky at the site of the observatory is very dark toward the North, where no constructions are located, and fair toward the South. The seeing varies a lot between nights, but can reach 1.2-1.5\arcsec on good nights.

The telescope was purchased from Torus, Inc., It is a 0.5m f/10 Cassegrain telescope, with an equatorial mount (see Fig \ref{fig1}). Its installation at the observatory was done in 2003. The theoretical pointing precision is about 0.1\arcsec \ for a field of view of 20\arcmin. The instrumentation is composed of a Marconi 42-40 back-illuminated CCD camera with 2048 $\times$ 2048 13.5$\mu$m pixels provided by Finger Lakes Instruments, and a 12 position filter wheel equipped with Johnson UBVRI and Clear filters. Both are computer controlled via an USB connection.

The weather control is performed by a Davis professional weather station, and a Boltwood cloud detector. Both are connected to the same computer which controls the dome door. The dome has been provided by Ash Domes, Inc., offering a fair rotation speed of 6 degrees per seconds. The dome has been tested to resist to the wind of two category 5 hurricanes, and is, thus, considered to be hurricane-proof.

To control the telescope, a program called {\it Talon} has been developed by Optical Mechanics, Inc. This program discusses with the control cards located on each axis of the telescope (HA, Dec, focus) to pilot the servomotors. These cards were developed by Clear Sky Institute, but are no longer supported. Technically, {\it Talon} allows for on-site or remote observations, and can follow a pre-defined observation schedule. In practice, due to the weather conditions during a typical Virgin Island night (frequent short showers in the start or middle of the night), it is safer to use it on-site only. For this reason, the facility also include two apartments for a resident astronomer and a visiting astronomer.

\section{Scientific objectives}
\label{objectif}

For about a decade, the science operations included: monitoring of optical counterparts of gamma-ray bursts \citep[GRB,][]{gib04}; coordinated observing campaigns of various objects; faculty and student research; and the educational/outreach missions of the university. This however led to few publications. A major step forward was done in 2015 with the definition of new scientific objectives that aimed at a better use of the facility and a larger number of published results.

\begin{figure}[!t]\centering
  \vspace{0pt}
  \includegraphics[width=0.9\columnwidth]{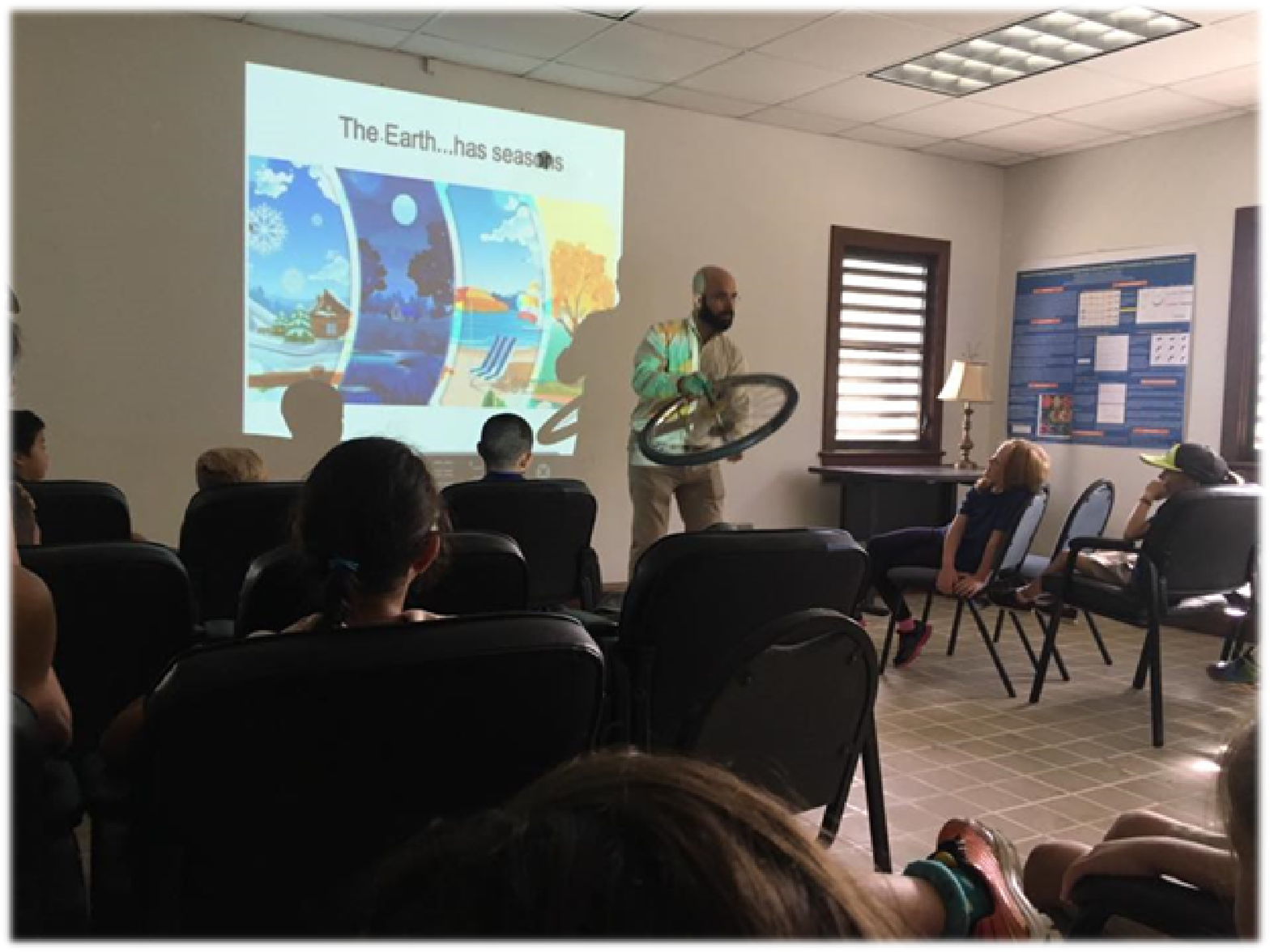}
	\includegraphics[width=0.9\columnwidth]{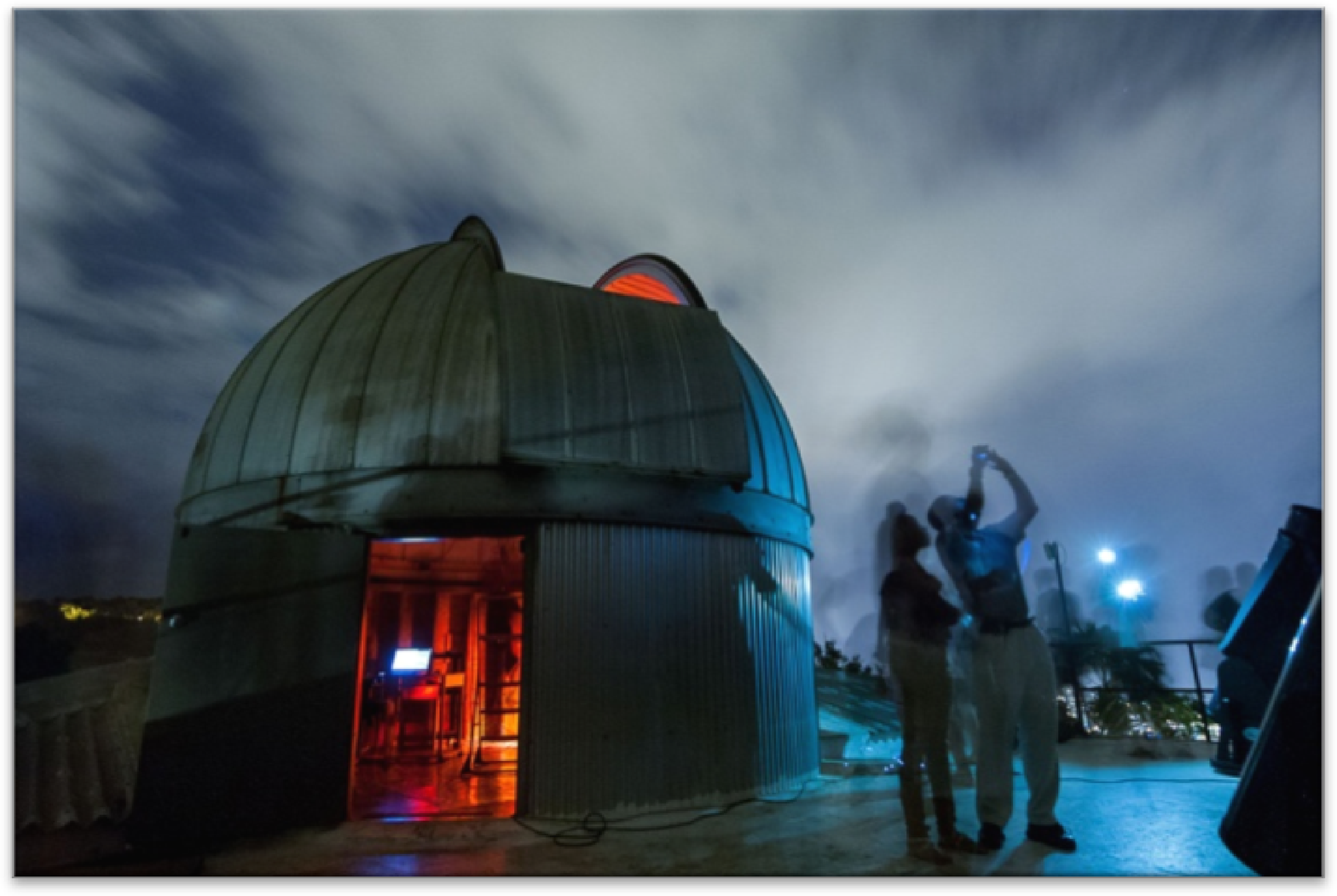}
  \caption{Examples of an outreach session at the facility. Top, a general presentation on a scientific fact (in that peculiar case, the rotation and revolution of the Earth). Bottom, a stargazing session on the roof of the observatory. The VIRT dome is opened to allow seeing the slews of the telescope.}
  \label{fig2}
\end{figure}

\begin{itemize}
\item[$\star$] Fast Transient Events	(30\% of available time). These events include GRBs, Fast Radio Bursts, Gravitational Wave counterparts, and various kinds of stellar outburst, which are visible only for a couple of minutes to hours, and with a short notice time. These events have also the priority on any other kind of observations.

\item[$\star$] Time Constrained Events	(20\% of available time). These events include occultation events (i.e. stars, solar system bodies, exoplanet transients), and predictable events, which can be scheduled in advance.

\item[$\star$] Slow Transient Events	(15\% of available time). These events are supernovae, novae, X-Ray binary outburst, tidal disruption events, and stellar flares that develop on several days and can suffer for a short delay in the observations.

\item[$\star$] Stellar Monitoring Campaigns	(15\% of available time). The advantage of this facility is that it is not heavily oversubscribed, and thus allows for repetitive observations for more than 6 months. This kind of science, now totally impossible on large instruments, is still well appreciated by stellar physicists. 

\item[$\star$] Space Debris Monitoring \& Sky Surveying	(10\% of available time).

\item[$\star$] Teaching	(5\% of available time). A small fraction of the observing time is devoted to astronomy labs, and student observing projects, as part of our University duties.

\item[$\star$] Outreach	(5\% of available time). We are also organizing public nights, where the observatory is open to anybody. During these nights, we are loading a special schedule asking for repetitive slews in order to explain to the audience the purpose of a robotic instrument and how it is working (see Fig. \ref{fig2}).
\end{itemize}

\section{How to reach the objectives? The ROS software}
\label{soft}

It appeared quickly once that new objectives were set that {\it Talon} was not suited for the tasks envisaged. Firstly, the computers installed in 2003 were too old for running a modern operating system. In fact, these computers were (in early 2015) running a version 6 of the Red Hat distribution, first made available in 1999, and totally outdated for both running new hardware and assuring a minimal security for an open system plugged to Internet. Additionally, {\it Talon} itself was not compatible with the large gap in libraries implied by the installation of new workstations running a Fedora Core distribution.

Some software updates were performed by OrangeWave Innovative Science LLC to allow resuming the activities of {\it Talon} on the new computers. It was however clear that {\it Talon} was lacking the short term reactivity needed to achieve the main scientific goal.

Thus, instead of fixing the old software, it was decided to install a new robotic system. We chose {\it ROS}, the Robotic Observatory Software developed by the TAROT team \citep{klo08}. The advantage of this code is that all the automation is already done and has been tested on various instruments. The only need was to develop a new library for interfacing the telescope commands into the ROS language. This was done in 2016, and tested in the early part of 2017.

ROS is versatile enough to allow for all the scientific goals to be fulfilled. In particular it has two main modes: a robotic mode, where observers request observations through a web interface and the system builds a dynamic schedule based on the last requests (an alert can always interrupt this schedule if needed), and an automated mode (called 'manual' in the ROS terminology) where the user bypass the schedule and send orders directly to the telescope through internet. In that latter case, the weather control is still operating and can close the roof if needed but the alerts are no longer processed. Both of these modes are made so that no astronomer has to be present on site to manage the observatory.

For some exceptional scientific cases, it is possible to pause the system and to use a full manual mode, running {\it Audela} \citep{klo12}; in such a case all automation is stopped and the weather control loop dismissed.

\section{Hardware replacements}
\label{hard}

The software update implied a recalibration of the pointing of the telescope, which shown a preoccupying aging of the telescope. The pointing accuracy for instance was more of the order of the degree than the sub-arcsecond. The telescope was also wobbling while tracking, preventing long exposures. Most of the first semester of 2017 was devoted to studying these features. 

\begin{figure}[!ht]\centering
  \vspace{0pt}
  \includegraphics[width=0.9\columnwidth]{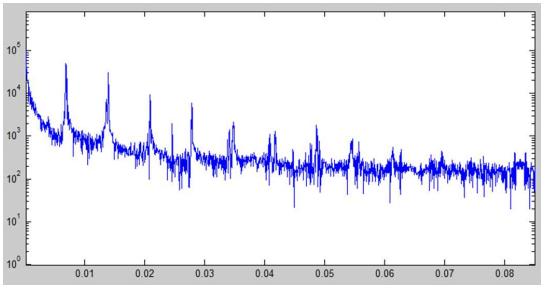}
  \caption{A fast fourier transformation of the measured offset between the requested pointing and the real value, while tracking. Some harmonics can clearly be seen at the frequencies expected from a problem of a small pulley inside the motor reducer.}
  \label{fig3}
\end{figure}

\begin{figure}[!ht]\centering
  \vspace{0pt}
  \includegraphics[width=0.9\columnwidth]{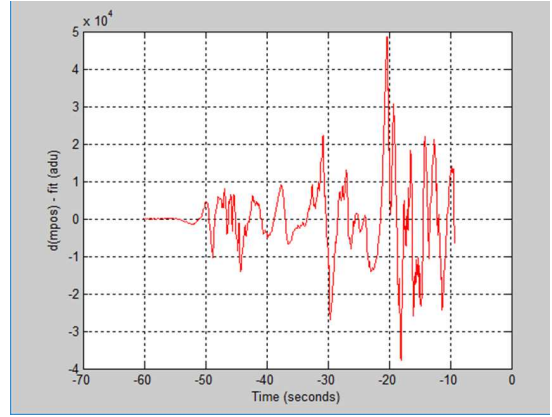}
  \caption{Offset between the requested position and the current observing position, while tracking. This is an extract from a longer test. The amplification of the wobbling is clearly seen in this diagram, and due to the tracking functions embedded in the CSIMC cards.}
  \label{fig4}
\end{figure}

\begin{figure*}[!htb]\centering
  \vspace{0pt}
  \includegraphics[width=1.9\columnwidth]{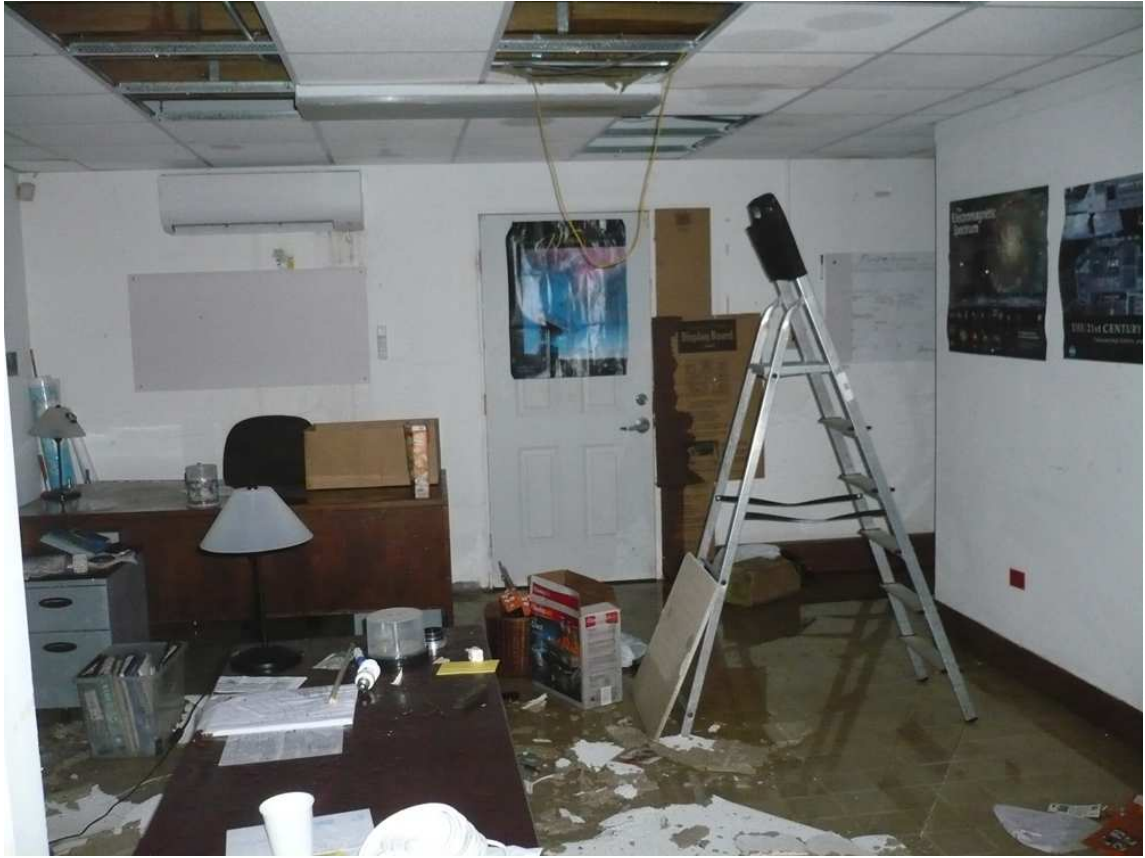}
  \caption{Picture of the control room taken after the hurricane Irma, and before the hurricane Maria. The computers were stored in a safe place; it is however clear from the picture a lot of work is due before the operations can resume at the Observatory.}
  \label{fig7}
\end{figure*}

It appeared that the optical encoders suffered from 10 years of use, as the motor and motor reducers (see Fig. \ref{fig3}) presented a cyclic pattern of drifting. We started by making some modifications to the friction wheel transferring the torque of the motors to the telescope on the HA axis, as it was damaged. This allowed to stop most of the drifts observed when the telescope was far from the meridian. 

A fine study of the tracking on several stars also indicated that one pulley in the reducer was damaged, inducing some wobbling. We also detected that the tracking functions embedded in the CSIMC cards, made to reduce any wobbling, had a tendency to increase it at contrary, due to the said malfunction of the pulley. This is illustrated in Fig. \ref{fig4}. We planed initially to replace these parts during the winter 2017 (see however the Conclusions).

Finally, we noted that the issue with the memory dump of the control cards was still present with the new system in place, without clear explanation. Several tests made on the cards indicated that the problem may be a memory leak from the pile, saturating the RAM and forcing a reboot of the card. Internet searches have shown that another team had the same issue, but it was not reported how they solved it (nor if it was possible). In a last attempt, we estimated that it was possible that the cards were corrupted by some corrosion, due to their age. We ran out of spares to test this hypothesis, and had to request new cards.

We then decided, at that point, to perform a main rebuilding of the telescope, and ordered a complete set of new motors, reducers, servomotors, control cards, and encoders. We will install these new elements as soon as we receive the shipments.

Finally, as the island is located in an active seismic area, we also performed a re-alignment of the telescope axis with the north pole, and discovered that the island slightly rotated during the last ten years, explaining the pointing inaccuracy.

\section{First results}
\label{result}

\begin{figure}[!hbt]\centering
  \vspace{0pt}
  \includegraphics[width=0.9\columnwidth]{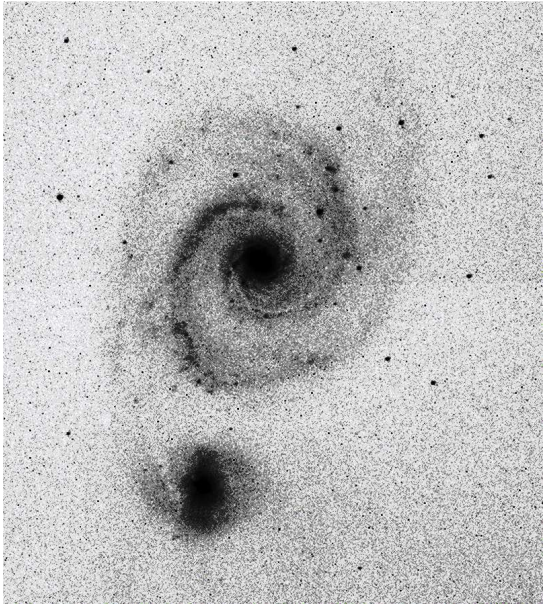}
  \caption{The whirlpool galaxy, taken as a calibration target for the optical path alignment. This image is a 30 seconds exposure in R band.}
  \label{fig5}
\end{figure}

\begin{figure}[!ht]\centering
  \vspace{0pt}
  \includegraphics[width=0.9\columnwidth]{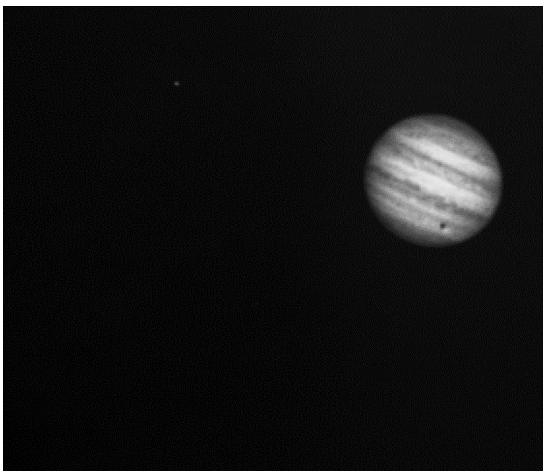}
  \caption{Jupiter, taken as a test of the dynamic of the camera. This is a 0.1 seconds exposure. It is possible to see one of the Galilean moons in the upper left corner of the picture.}
  \label{fig6}
\end{figure}

Despite being highly impacted by the hurricanes Irma and Maria (see Fig. \ref{fig7} and below), the observatory had performed several observations (see examples in Fig. \ref{fig5} and \ref{fig6}), some of them leading to publications. They are extensively described in Morris et al. (these proceedings), we only summarize them here.

We performed the follow-up of three gamma-ray bursts (GRB 170202A, GRB 170208B, and GRB 170405A). Two of these observations led to upper limits \citep{gen17b, gen17c}; however the observation of GRB 170202A provided a detection and several measurements of the light curve in the R band \citep{gen17a}. In addition, we participated on the observation campaign of GW170817 \citep{abb17, gen17d}. Being observed right before the September storms, the data were sent to Australia, where colleagues performed the data reduction and analysis, leading to a joint publication \citep{and17}.

\section{Conclusions}

The robotization of the Virgin Islands Robotic Telescope is now in its final stage. It was due for the second semester 2017. However, in September 2017, two hurricanes and the outer parts of two other storms (Harvey and Jose) hit St Thomas, leading to strong damages on the islands. Since then (to February 2018), the observatory is without power or internet, and all orders and deliveries have been canceled. 

The observations performed before these storms have confirmed the potential of the observatory and its complementarity with other instruments. Thus, despite the damages, we are still committed to complete this task. ROS allows for some kind of resilience, being modular. No data has been lost, and the telescope itself has not suffered any apparent damage. As soon as the power is restored in our area of the island, it should be possible to start operating the telescope again.

We will finish to install the new hardware during the Spring 2018, and we plan to test the new settings in early April-May. This should allow for a full robotic usage of the observatory during the summer, in line with the first public run of LIGO/Virgo. 2018 would then set, with the aftermaths of the hurricanes Irma and Maria, the final stone of a path opened by another hurricane, 20 years ago.

\adjustfinalcols

%\begin{figure}[!t]\centering
  %\vspace{0pt}
  %\includegraphics[angle=-90,width=0.8\columnwidth]{example-fig}
  %\caption{example}
  %\label{fig1}
%\end{figure}
%
%
%
%\begin{table}[!t]\centering
  %\setlength{\tabnotewidth}{\columnwidth}
  %\tablecols{3}
  %% Stretch the space between table columns 
  %\setlength{\tabcolsep}{2.8\tabcolsep}
  %\caption{A Simple Table (\lowercase{$x=1.0$})\tabnotemark{a}} \label{tab:ion_ab}
  %\begin{tabular}{lrr}
    %\toprule
    %Ion & \multicolumn{1}{c}{NGC~5461} & \multicolumn{1}{c}{NGC~5471} \\
    %\midrule
    %O$^0$    & $7.08\pm0.20$ & $6.63\pm0.20$\\
    %O$^+$    & $8.08\pm0.14$ & $7.32\pm0.14$\\
    %O$^{++}$ & $8.32\pm0.07$ & $8.02\pm0.07$\\
    %N$^+$    & $7.04\pm0.12$ & $6.01\pm0.13$\\
    %Ne$^{++}$ & $7.59\pm0.11$ & $7.32\pm0.10$\\
    %S$^+$    & $6.02\pm0.19$ & $5.47\pm0.20$\\
    %S$^{++}$ & $7.00\pm0.10$ & $6.45\pm0.10$\\
    %Cl$^{++}$ & $4.93\pm0.16$ & $4.20\pm0.16$\\
    %Ar$^{++}$ & $6.15\pm0.12$ & $5.55\pm0.14$\\
    %Ar$^{3+}$ &\multicolumn{1}{c}{\nodata} & $5.07\pm0.10$\\
    %\bottomrule
    %\tabnotetext{a}{Note the use of \CS{lowercase} to prevent the $x$
        %from being converted to upper case.}
  %\end{tabular}
%\end{table}


\begin{thebibliography}
\bibitem[Abbott et al.(2017)]{abb17} Abbott, B.P., Abbott, R, Abbott, T.D., et al.\ 2017 ApJL 848, L12
\bibitem[Andreoni et al.(2017)]{and17} Andreoni, I., Ackley, K., Cooke, J., et al.\ 2017, PASA, 
\bibitem[Neff et al.(2004)]{nef04} Neff, J.E., Allen, D.K., Aurin, D.M., Boyajian, T.S., Crowther, P., David, K., Drost, D.M., Giblin, T.W., Hurley, A., Lucas, J., Nations, H., Smith, D., Thomas, N., \& Walsh, M\ 2004, AN, 325, 669
\bibitem[Gendre et al.(2017a)]{gen17a} Gendre, B., Orange, N.B., Morris, D.C., et al.\ 2017a, GCN, 20598, 1
\bibitem[Gendre et al.(2017b)]{gen17b} Gendre, B., Orange, N.B., Morris, D.C., et al.\ 2017b, GCN, 20636, 1
\bibitem[Gendre et al.(2017c)]{gen17c} Gendre, B., Orange, N.B., Morris, D.C., et al.\ 2017c, GCN, 20998, 1
\bibitem[Gendre et al.(2017d)]{gen17d} Gendre, B., Cucchiara, A., Morris, D.C., et al.\ 2017d, GCN, 21609, 1
\bibitem[Giblin et al.(2004)]{gib04} Giblin, T.W., Neff, J.E., Hakkila, J., David, K., \& Hartmann, D.\ 2004, AN, 325, 670
\bibitem[Klotz et al.(2008)]{klo08} Klotz, A., Bo\"er, M., Eysseric, J., et al.\ 2008 PASP 120, 1298
\bibitem[Klotz et al.(2012)]{klo12} Klotz, A., Delmas, R., Marchais, D., Pujol, M., \& Jasinski, C.\ 2012 ASI Conference Series 7, 15
\end{thebibliography}
\end{document}